\documentclass[aps,prb,twocolumn,notitlepage]{revtex4-2}
\usepackage{epsfig,subfigure,amsmath,float,xcolor,ulem,physics,bm}
\usepackage[colorlinks]{hyperref}
\hypersetup{colorlinks=true,linkcolor=blue,filecolor=blue,urlcolor=blue,citecolor=blue}
\usepackage[utf8]{inputenc}
\usepackage{amssymb}
\usepackage{svg}
\usepackage{soul}
\newcommand{\beq}{\begin{equation}}
\newcommand{\eeq}{\end{equation}}
\newcommand{\bea}{\begin{eqnarray}}
\newcommand{\eea}{\end{eqnarray}}
\newcommand{\non}{\nonumber}
\newcommand{\si}{\sigma}
\newcommand{\al}{\alpha}
\newcommand{\om}{\omega}
\newcommand{\Om}{\Omega}
\newcommand{\eps}{\epsilon}
\newcommand{\beps}{\bar{\epsilon}}
\newcommand{\ig}{\includegraphics}
\newcommand{\hH}{\hat{H}}
\newcommand{\mK}{\mathbf{K}}

\begin{abstract}
A parametrically driven classical harmonic oscillator exhibits resonant instability when driven at twice its natural frequency, with the lowest energy configuration remaining unaffected by the drive. In contrast, the ground state of the quantum mechanical counterpart shows a non-trivial response to such a drive due to the spatial delocalization of the wavefunction. The standard realization of PR involves modulating the natural frequency of the oscillator. Here we study a different drive protocol in which the coupling between two such quantum harmonic oscillators is modulated parametrically. We show that the drive frequency can in principle be tuned to selectively excite any desired normal mode, while leaving the other close to its ground state. Only states with even quantum numbers in each normal mode are populated.  Within the parametric resonance window the excitations follow a power-law decay with respect to occupation number, in contrast to the exponential decay observed off-resonance. We also briefly discuss how this framework can be extended to a system of $N$ coupled  oscillators.
\end{abstract}

\begin{document}

\title{Mode-selective excitation in parametrically driven coupled quantum oscillators}

\author{Ranjani Seshadri}
\email{ranjani.seshadri@christuniversity.in}
\email{ranjani.physics@gmail.com}
\affiliation{Department of Physics and Electronics, Christ University, Bengaluru, India, 560029}
\date{\today}
\maketitle

\section{Introduction}
Periodically driven quantum systems have emerged as a rich playground for exploring novel physics \cite{gold14, oka09, sesh19, sesh22, lin11}. Such systems, which can exhibit emergent properties absent in the equilibrium counterpart, have attracted significant attention across multiple fields such as condensed matter physics, quantum optics and quantum information. A particularly elegant way to drive a quantum system is by making a parameter of the Hamiltonian time-dependent thereby modulating the potential landscape directly. This is unlike ordinary resonance, where an external force directly couples to a coordinate of the system.

A classical example of such a driven system is the textbook problem of parametric resonance (PR), where the natural frequency of a harmonic oscillator is made time-dependent in a certain way. The most natural physical example of this phenomenon is when the suspension point of a simple pendulum is oscillated vertically. This well studied classical problem \cite{landau,safonov19,dyk06,kawalec17} is described by the following equation of motion,
\beq
\ddot y  + \omega_0^2f(t)y=0 \label{eq:HO}
\eeq
with $f(t) = (1+h \sin(({2 \om_0+\eps})t)$. Here, $h$ and $\eps$ are the amplitude and the detuning parameter, respectively. For the classical oscillator it is a well known result that PR occurs when $|\eps|<\frac{1}{2}h\omega_0$ \cite{landau}.

When the same parametric drive is applied to a quantum harmonic oscillator, the same condition holds for PR to occur. However, the primary difference between the classical and quantum oscillators is that while the classical system in its lowest energy configuration does not respond at all to this drive, the quantum system shows excitations even if it is in the ground state to begin with \cite{sesh25}. 

{Parametrically driven quantum oscillators arise in a wide range of contexts in physics \cite{bone2024,dyk11}. Superconducting quantum circuits use Josephson parametric amplifiers to amplify weak microwave signals \cite{marth18,wust13}. In quantum opto-mechanics \cite{garcia25, cart18, nguyen21, han24, calvo99, Lu_23, lifshitz2003} parametric driving is used to generate entanglement between optical and mechanical modes. Parametric driving is also relevant in the context of ion traps {and cavity} resonators \cite{vasant92, clem18}. {Yet} another striking application occurs in the context of quantum field theory when one can  parametrically excite electromagnetic fields \cite{bran,qprfield}. In all these systems, a time dependence is introduced by varying (with time) a specific parameter in the Hamiltonian. One of the most commonly studied realizations of this is when the natural frequency of the system is made time dependent, i.e., a single parametrically driven quantum oscillator \cite{weig2002,akridge95,grub19,shao08}.}

In this work, we extend the study of parametric resonance to a system of coupled quantum harmonic oscillators. Parametric driving is incorporated by modulating the coupling between the two degrees of freedom rather than the natural frequency. Although the analysis and numerical results that follow are for the two-oscillators case, we will later argue that it can be generalized to $N$ coupled oscillators. This paper is organized as follows. We start in Sec. \ref{sec:TDSE} with the normal-mode description of a two-dimensional quantum harmonic oscillator with a coupling term modulated in the form of a parametric drive. This is followed in Sec. \ref{sec:Numevol} by a discussion of the time-evolution of the ground state of such a system. The numerical solution hinges on the fact that a Gaussian wave packet remains a Gaussian under a quadratic Hamiltonian. We then derive selection rules which restrict the excitations to even quantum numbers in each normal mode. In Sec. \ref{sec:Results} we present the main numerical results, organized around two distinct regimes: vanishing static coupling, where the two normal modes are degenerate and get excited symmetrically, and a finite static coupling, where the degeneracy is lifted and the drive frequency can be tuned to selectively excite a single mode while leaving the other close to its ground state. This is followed by a brief discussion on the generalization to $N$ coupled oscillators in Sec. \ref{sec:Nosc}. We conclude in Sec.~\ref{sec:disc} with a summary of the results and a discussion of possible applications and extensions.

\section{Quantum Oscillator with time-dependent Coupling} \label{sec:TDSE}

We consider a particle of mass $m$ confined in a two-dimensional isotropic harmonic potential with a time-dependent mode coupling {parameterized by a function $\kappa(t)$. The 
dynamics of this particle is governed by the time-dependent Schr\"odinger equation,
\beq
i\hbar \frac{\partial}{\partial t}\phi(x,y,t)=\hH\phi(x,y,t).
\label{TDSE_xy}
\eeq
where the Hamiltonian is given by,
\beq
\hH=-\frac{\hbar^2}{2m}\left(\frac{\partial^2}{\partial x^2}+\frac{\partial^2}{\partial y^2}\right)+\frac{1}{2} m \om_0^2 (x^2 + y^2)+\kappa(t)\, x y .
\label{H_xy}
\eeq
Although this Hamiltonian is written in terms of two spatial coordinates, it more generally describes any quantum harmonic oscillator with two degrees of freedom coupled through such an interaction term.

Introducing dimensionless scaled coordinates $\xi = \al x$ and  $\eta = \al y$ with $\al = \sqrt{m \om_0/\hbar}$ and dimensionless time $\tau = \om_0 t$, the {(dimensionless)} Hamiltonian, $\hH / (\hbar \omega_0)$ is expressed as,
\begin{align}
i\frac{\partial}{\partial\tau}=-\frac{1}{2}\left(\frac{\partial^2}{\partial \xi^2}+
\frac{\partial^2}{\partial \eta^2}\right)+\frac{1}{2}
(\xi^2 + \eta^2)+{\beta(\tau)}\, \xi \eta,
\label{H_xieta}
\end{align}
with $\beta(\tau)\equiv \kappa(t)/(m\om_0^2)$.
We then transform to normal mode coordinates $(Q_+,Q_-)$ which are related to $(\xi,\eta)$ by a $\pi/4$ rotation, i.e.
\beq
Q_\pm = \frac{\xi\pm\eta}{\sqrt{2}}. \label{eq:xieta}
\eeq
These are the symmetric (in-phase,$+$) and antisymmetric (out-of-phase, $-$) combinations of $\xi$ and $\eta$. This decomposition is exact for any isotropic potential with a bilinear coupling of the form $\kappa(t)\, xy$.

The time-dependent Schr\"odinger equation (TDSE) in these coordinates reads
\begin{subequations}
\beq
i\frac{\partial \Psi}{\partial \tau} = \left(\hat H_+(\tau) + \hat H_-(\tau)\right) \Psi,
\eeq
where $\Psi(Q_+,Q_-,\tau)\equiv\phi(x,y,t)$ and,
\beq
\hat H_\pm = -\frac{1}{2}\frac{\partial^2}{\partial Q_\pm^2} + \frac{1}{2}\,\Omega_\pm^2(\tau)\,Q_\pm^2.
\label{eq:Hnorm}
\eeq
The ``dimensionless natural frequencies" of the normal modes are given by
\beq
\Om_\pm^2(\tau)=1 \pm{\beta(\tau)}.
\eeq
\end{subequations}
Let us now assume a specific time dependence of the original mode coupling,
\bea 
\kappa(t) = \begin{cases} \kappa_0 +\kappa_1\sin[(2\om_0+\eps)t]~~~~~ \text{if}~~~~ 0<t<t_f \\ \kappa_0, ~~~~~~~~~~~~~~~~~~~~~~~~~~~~\text{otherwise.}\end{cases} 
\eea
 Here, the coupling $\kappa(t)$ is modulated at frequency $\omega = 2\omega_0 + \eps$,  where $\eps$ is a small detuning parameter. The drive is applied for a finite duration $0 < t < t_f$, corresponding to an 
integer number of cycles, after which the coupling returns to its 
static value $\kappa_0$.
\begin{figure}[htb]
    \centering
    \ig[width=0.8\linewidth]{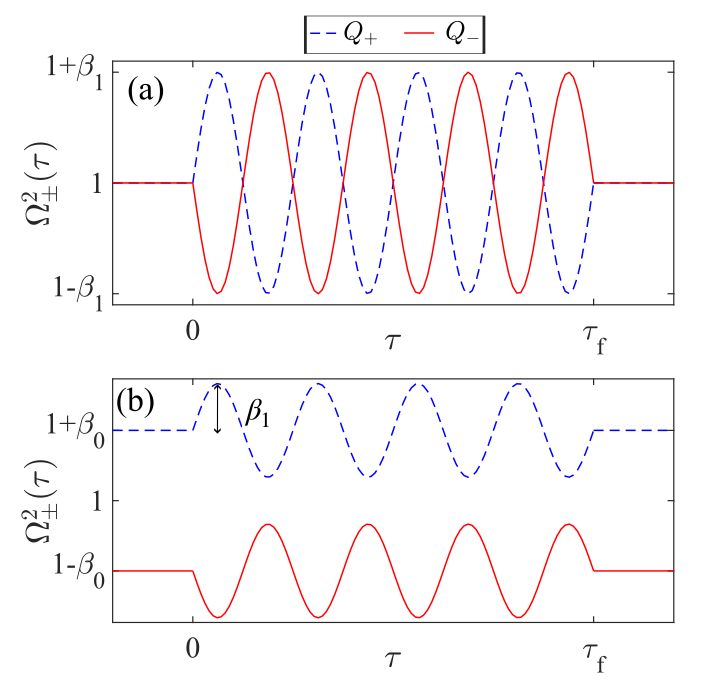}
    \caption{The time dependence of the normal mode frequencies (dimensionless) for the two cases (a) $\beta_0 = 0$ and (b) $\beta_0\neq0$. In the former the  oscillation is about the same mean value while in the latter it is about two different mean values. The amplitude of oscillations $=\beta_1$ is same in both the cases.}
     \label{fig:omegapm}
\end{figure}

This means the normal-mode coordinates have time-dependent ``natural" frequency given by
\begin{subequations}
\bea 
\Om_\pm^2(\tau) = \begin{cases} 1 \pm\beta_0\pm \beta_1\sin[(2+\bar\eps)\tau],~~~ \text{if}~ 0<\tau<\tau_f \\ 1\pm\beta_0, ~~~~~~~~~~~~~~~~~~~~~~~~\text{otherwise}\end{cases}
\eea
with $\beta_{0(1)} = \kappa_{0(1)}/m \om_0^2$ being the dimensionless coupling parameter, and $\bar{\eps} = \eps/\omega_0$ is the dimensionless detuning parameter, measuring the fractional deviation of the drive frequency $\omega$ from the resonance condition $\omega = 2\omega_0$. In the standard form, this becomes:
\bea 
\Om_\pm^2(\tau) = \begin{cases} \Om^2_{\pm0}(1 \pm h_\pm \sin[(2+\bar\eps)\tau],~~\text{if}~ 0<\tau<\tau_f \\ {\Om^2_{\pm0}}, ~~~~~~~~~~~~~~~~~~~~~~~~~~\text{otherwise}\end{cases}
\eea
\label{eq:defmodes}
\end{subequations}
with $\Om_{\pm0} = \sqrt{1\pm\beta_0}$ and $h_\pm = \beta_1/\Om^2_{\pm0}$. Each normal mode, therefore, realizes an independent parametrically driven harmonic
oscillator. The two modes experience out-of-phase modulation. This behavior is shown in Fig. \ref{fig:omegapm} for two cases (a) $\beta_0=0$ where the two frequencies oscillate about the same mean value and (b) $\beta_0 \neq 0$ where they oscillate about two different mean values. The driving of the second kind shown in Fig. \ref{fig:omegapm}(b) is what results in the mode selective behavior.

\section{Evolution of Ground state} \label{sec:Numevol}

\subsection{Ground state at Equilibrium}
Before the drive is switched on ($\tau<0$), the system is prepared in the ground state of the two-dimensional harmonic oscillator. Since the sHamiltonian separates into normal-mode coordinates, the full wave function factorizes as,

\bea
\Psi(Q_+,Q_-,\tau)={\braket{Q_+,Q_-}{\Psi}}_\tau = \prod_{\si=\pm}\psi_\si(Q_\si,\tau).
\eea

Accordingly, the ground state $\ket{0_+,0_-}$, is a separable Gaussian,
\begin{subequations}
\bea
{\braket{Q_+,Q_-}{0_+,0_-}}_{\tau<0}
= \prod_{\si=\pm} \braket{Q_\si}{0_\si} \non \\
= \prod_{\si=\pm} \left(\frac{\Om_{\si 0}}{\pi}\right)^{\frac{1}{4}}
\exp\left(-{\tilde Q_\si^2}/{2}\right),
\eea
with the zero point energy
\beq
E_{0_+,0_-} = \frac{1}{2}\hbar \om_0 \left(\Om_{+0}+\Om_{-0}\right).
\eeq
where $\tilde Q_\si = \sqrt \Om_{\si0} Q_\si$. 
\end{subequations}
More generally, any energy eigenstate $\ket{n_+,n_-}$ has energy,
\begin{subequations}
\beq
E_{n_+,n_-}=\hbar\om_0 \left[\Omega_{+0} \left(n_+ + \frac{1}{2}\right) + \Omega_{-0}\left(n_- + \frac{1}{2}\right)\right]
\eeq
and wavefunction
\bea
&~&\braket{Q_+,Q_-}{n_+,n_-} = \prod_{\si=\pm} \braket{Q_\si}{n_\si} \non \\
&=& \prod_{\si=\pm} \left(\frac{\Om_{\si 0}}{\pi}\right)^{\frac{1}{4}}
\frac{1}{\sqrt{2^{n_\si} n_\si !}} H_{n_\si}\!\left(\tilde Q_\si\right)
\exp\!\left(-{\tilde Q_\si^2}/{2}\right).\non \\
\label{eq:En}
\eea 
\end{subequations}

\subsection{Time evolution of the ground state}

We now move on to analyzing the time-evolution of the quantum ground state under this parametric drive in terms of the normal mode coordinates. Since a Gaussian wave-function under a quadratic Hamiltonian continues to remain a Gaussian with a different set of parameters, we can write the general time-dependent wave function for each normal mode $\sigma$ as,

\bea
\psi_{\si}(Q_\si,\tau) = \begin{cases}  
A_\si(\tau)e^{-B_\si(\tau)Q_\si^2} \quad \quad \text{if    } 0<\tau<\tau_f \\ \\
\left(\frac{\Om_{\si0}}{\pi}\right)^{1/4}
e^{-\Om_{\si0}Q_\si^2/2} \quad \quad \text{if    } \tau<0  \end{cases} \label{eq:psitau}
\eea
Note that the full wave function at any instant is the product of $\psi_\si$ for $\si=\pm$. Additionally, the time-dependent coefficients $A_\si$ and $B_\si$ are allowed to be complex. By substituting Eq.~\eqref{eq:psitau} into the TDSE Eq.~\eqref{eq:Hnorm}, and comparing the powers of $Q_\si$, we get four coupled differential equations for each mode $\si$,
\bea
A_{\si,R} B_{\si,R} - A_{\si,I} B_{\si,I} + \dot{A}_{\si,I} &=& 0 \non \\ 
A_{\si,R} B_{\si,I} + A_{\si,I} B_{\si,R} - \dot{A}_{\si,R} &=& 0 \non \\ 
\Omega^2_\sigma(\tau)-2\dot{B}_{\si,I}-4(B_{\si,R}^2-B_{\si,I}^2) &=& 0 \non \\
\dot{B}_{\si,R} - 4 B_{\si,R} B_{\si,I} &=& 0, 
\label{eq:diff}
\eea
with the subscripts $R$ and $I$ denoting the real and imaginary parts of these quantities. Numerically solving these gives us the evolved state $\ket{\Psi(\tau_f)}$ at the end of the drive. This has components not only in the unperturbed ground state but also in the excited states, because, while it is still a Gaussian, it now has a set of parameters different from that of the ground state. Therefore, the probability $p_{n_+,n_-}$ of finding the evolved state in $\ket{n_+,n_-}$ is,
\beq
p_{n_+,n_-} = |\braket{n_+,n_-}{\Psi}|^2 =  \left|\braket{n_+}{\psi_+}\braket{n_-}{\psi_-}\right|^2 
\label{eq:projection}
\eeq
where we have written the evolved state as $\ket \Psi$ for brevity. The numerical analysis is very similar to the case of a single parametrically driven harmonic oscillator discussed in \cite{sesh25}.

\subsection{Selection Rules}
The structure of the evolved state imposes strong constraints on the allowed transitions with a transparent physical origin. Since the full wavefunction factorizes as $\Psi = \psi_+\psi_-$ at all times, the overlap in Eq.~\eqref{eq:projection} splits into a product of independent overlaps for each normal mode. The parametric drive preserves the even parity of $\psi_\sigma(Q_\sigma)$ under $Q_\sigma \to -Q_\sigma$ for each $\sigma = \pm$ separately, while the harmonic oscillator eigenstate $|n_\sigma\rangle$ has definite parity $(-1)^{n_\sigma}$. Therefore the overlap 
$\braket{n_\sigma}{\psi_\sigma}$ vanishes identically for all odd $n_\sigma$ in \textit{either} mode independently, i.e.,
\beq
p^{(\sigma)}_{n_\sigma} = 0 \quad \text{for odd } n_\sigma,
\eeq
Hence, $p_{n_+,n_-} = 0$ unless both $n_+$ and $n_-$ are 
even integers. The system therefore climbs the ladder of each 
normal mode two rungs at a time. Note that this constraint is 
stronger than requiring even total excitation number $N = n_+ + n_-$: states such as $|1,1\rangle$, which have $N=2$ but odd quantum number in each mode individually, are nonetheless forbidden.

\section{Numerical results} \label{sec:Results}
The parameters $\beta_0$ and $\beta_1$ correspond to the mean value of coupling and the amplitude of the drive respectively. The analysis of the system is carried out separately for the two distinct regimes: $\beta_0 = 0$ and $\beta_0 \neq 0$, corresponding, respectively, to vanishing and finite static mode coupling as can be inferred from Eq. \ref{eq:defmodes}. The main results are as follows.

\subsection{$\beta_0=0$}
In this case the two normal modes are degenerate, with $\Om_{+0} = \Om_{-0} = 1$ and therefore the zero-point energy is simply $E_{0+,0-} = \hbar\om_0$. As a consequence, both modes share the same PR condition and are excited identically under the drive. This is evident in Fig.~\ref{fig:pmn_b0}, which shows the probability $p_{n_+,n_-}$ as a function of $\beps$ for various states $(n_+,n_-)$. We observe that the behavior of $p_{n_+,n_-}$ shows a fairly sharp transition at $\beps = \pm \beta_1/2$. Outside this window the probability of excitation to higher levels is very small and $p_{00}$ close to $1$. This means that the parametric drive does not significantly excite the higher energy states and the system remains close to the ground state provided $|\beps| > \beta_1/2$.

\begin{figure}[htb]
    \centering
    \ig[width=0.78\linewidth]{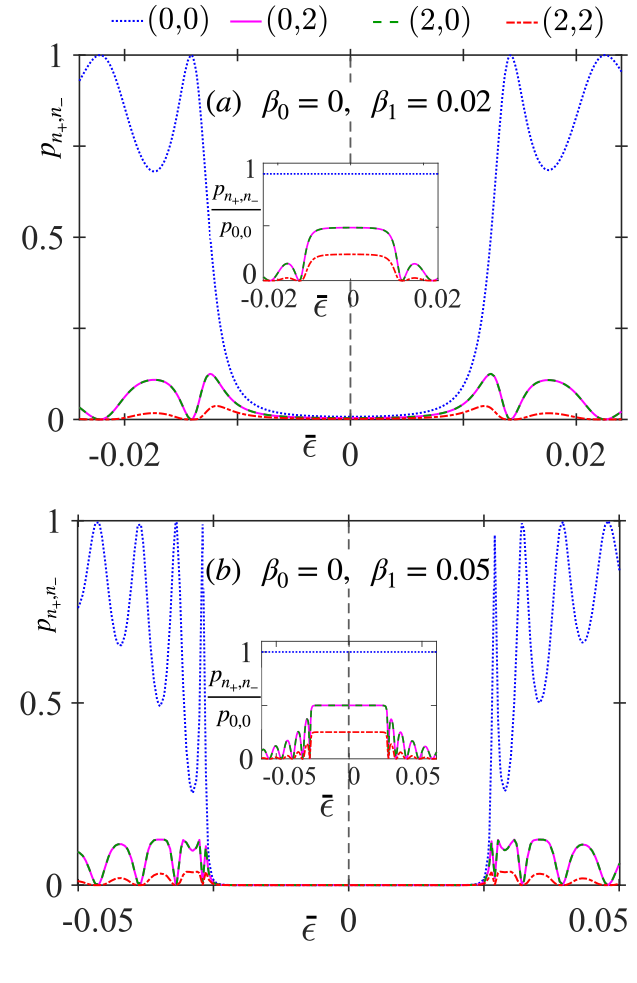}
    \caption{Probability $p_{n_+,n_-}$ for parameters (a) $\beta_0 = 0,~\beta_1 = 0.02$, and (b) $\beta_0 = 0~,\beta_1 = 0.05$. The different curves are for the different states as shown in the legend. The number of drive cycles is $\nu=200$. Since the resonance frequency is same for both modes, the states $(0,2)$ and $(2,0)$ behave identically. The PR window is approximately $\beps \in [-\beta_1/2, \beta_1/2]$. Outside this range the higher energy states are not significantly occupied. The insets show the relative probabilities for the excited states normalized with respect to the ground state $(0,0)$.}
    \label{fig:pmn_b0}
\end{figure}

 The behavior within this window requires a more careful analysis. The small value of $p_{00}$ means that the higher energy states are being populated. But we also note that the {\it individual} values of $p_{n_+,n_-}$ for the higher states is also small. This indicates that the energy is transferred (and hence gets distributed) to several excited states. As a result, the individual values of $p_{n_{+}n_{-}}$ including for $p_{00}$ are small. To understand the nature of $p_{n_{+}n_{-}}$ in the resonance window, it is useful to look at the ratio $p_{n_{+}n_{-}}/p_{00}$ which we have plotted in the inset of Fig. \ref{fig:pmn_b0}. It shows that in the resonance window, $p_{n_{+}n_{-}}$ for $(n_+,n_-) \ne (0,0)$ is a substantial fraction of 
$p_{00}$, indicating significant excitations. 

Furthermore, due to the degeneracy of the modes, states related by exchange of quantum numbers—such as $(0,2)$ and $(2,0)$—exhibit identical population dynamics. This is evident from the overlapping magenta and green curves in both the panels in Fig. \ref{fig:pmn_b0}. Note that for finite-time driving, the boundary between the two regimes is not abrupt and becomes smooth instead. The behavior of individual normal-modes is similar to the one-dimensional case discussed in Ref.~\cite{sesh25}.

To better visualize the structure of excitations in the two-dimensional Hilbert space, we plot the probability distribution $p_{n_+,n_-}$ as a function of $(n_+,n_-)$ in Fig. \ref{fig:pmn_surf}. In (a), i.e. outside the PR regime, we see that only a very few states close to 0 in each of the modes are occupied. In contrast, in (b), i.e. within the PR regime  a significant number of higher energy states are excited in both the normal modes. We also see that there is a reflection symmetry under $n_+ \leftrightarrow n_-$ indicating that the normal modes are at the same footing which is a result of the degeneracy between the two modes.

\begin{figure}[htb]
\ig[width = 0.98\linewidth]{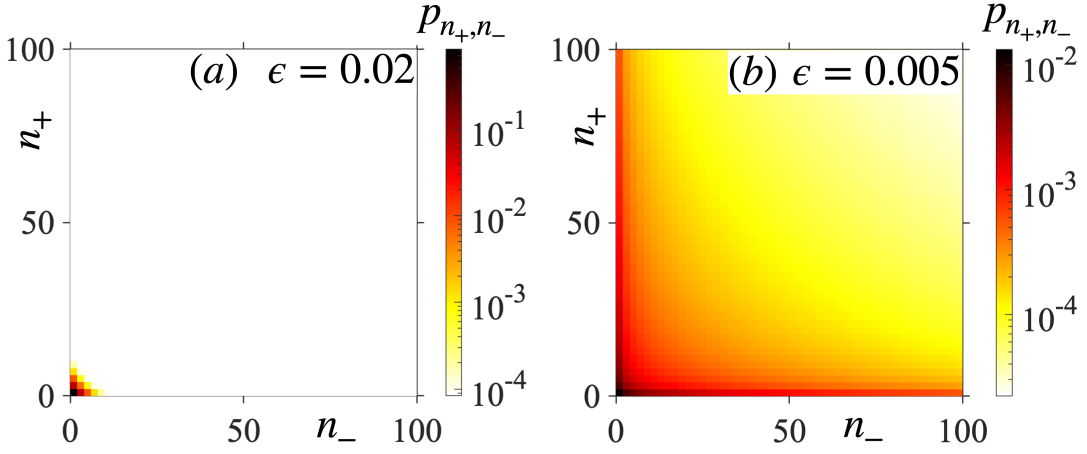}
\caption{ Color plots of the excitation probabilities $p_{n_+,n_-}$ in the normal-mode basis for two representative values of the detuning parameter $\eps$. (a) For $\eps = 0.02$, outside the PR regime, the distribution is strongly localized near the ground state, indicating negligible excitation. (b) For $\eps = 0.005$, within the resonance window, the distribution spreads over a wide range of $(n_+,n_-)$, reflecting significant excitation of higher-energy states. The color scale is logarithmic.}
\label{fig:pmn_surf}
\end{figure}

\subsection{$\beta_0\neq0$}
In this case, the degeneracy of the two normal modes is lifted, leading to distinct frequencies $\Omega_{+0} \neq \Omega_{-0}$. Therefore, the PR condition is no longer identical for the two modes, and instead splits into two separate resonance windows given by $\beps_\pm \in [\beta_0-\beta_1/2, \beta_0+\beta_1/2]$ approximately as shown in Fig. \ref{fig:pmn_b1}. We have shown the behavior for two different sets of drive parameters (a) $\beta_0 = 0.05,~ \beta_1 = 0.01$ and (b) $\beta_0 = 0.05,~ \beta_1 = 0.02$. In both these since $\beta_0$ is the same, the centers of the resonance windows remain the same. However, due to the change in the value of $\beta_1$, the widths of the PR windows are different. Another difference here from the $\beta_0=0$ case is that the curves corresponding to $(0,2)$ and $(2,0)$ are no longer overlapping. From the inset we can see that the amplification in the probabilities for these modes occur in two different ranges of $\beps$. This has important implications for the dynamics of this driven system. For a given value of the detuning parameter $\beps$, only one of the two modes can be brought close to resonance, while the other remains predominantly in the ground state. As a result, the excitation becomes strongly mode-selective, in contrast to the symmetric excitation observed in the degenerate case $\beta_0 = 0$.

\begin{figure}[htb]
    \centering
    \ig[width=0.78\linewidth]{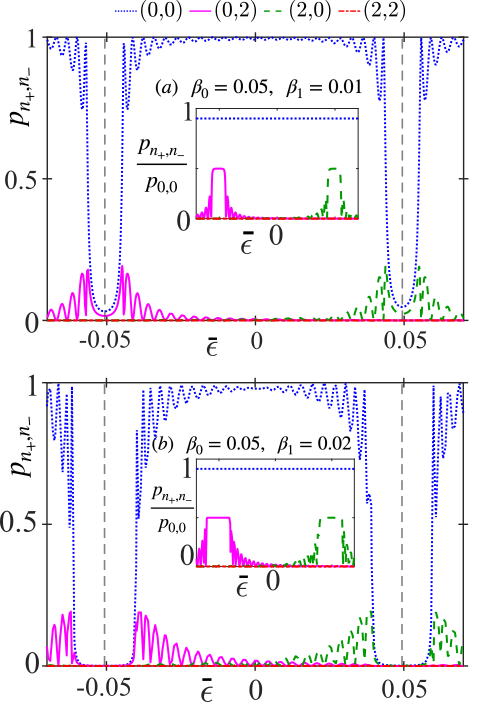}
    \caption{Probability $p_{n_+,n_-} = |\braket{n_+,n_-}{\Psi}|^2$ for drive parameters. The different curves in each plot are for the states $(0,0)$ (blue dotted), $(0,2)$ (magenta solid),$(2,0)$ (green dashed) and $(2,2)$ (red dash dot). The number of drive cycles $\nu=500$ for all the panels. Since $\beta_0\neq 0$,  the two normal modes are non degenerate. Therefore the states $(0,2)$ and $(2,0)$ are now distinct in contrast with the $\beta_0=0$ case. The presence of a non-zero $\beta_0$ now creates two separate windows where the two normal modes undergo PR. The width of the resonance windows is different in (a) and (b) due to different values of the drive amplitude $\beta_1$. The insets show the relative probabilities for the excited states normalized with respect to the ground state $(0,0)$. We clearly see that in both cases, within the PR window $\beps \in [\beta_0-\beta_1/2, \beta_0+\beta_1/2]$ (approximately), the higher states get more excited as compared to outside this range of $\eps$.}
    \label{fig:pmn_b1}
\end{figure}

Comparing Figs. \ref{fig:pmn_b1}(a) and \ref{fig:pmn_b1}(b) it is also evident that for a given $\beta_0$ the separation of the two windows decreases with increasing $\beta_1$. This means that mode-selective excitation is possible only as long as there is no overlap between these windows, i.e. $\beta_1 \lesssim  2\beta_0$.

\subsection{Population of excited states} \label{sec:sspop}

To characterize the structure of the excitation spectrum when $\beta_0=0$, we examine the average probability per state, $P_N/g_N$, as a function of the total occupation number $N = n_+ + n_-$, where $g_N=\frac{N}{2}+1$ is the degeneracy  of the $N$-th level counting only the states allowed by the selection rule. This quantity captures how probability is distributed across energy shells in the two-dimensional Hilbert space.

Within the PR window, the distribution exhibits a power-law decay,
\begin{equation}
    \frac{P_N}{g_N} \propto N^{-\al_{\text{pow}}},
    \label{eq:pow}
\end{equation}
with $\al_{\text{pow}}>0$. This implies a broad, heavy-tailed distribution of excitations: higher energy states have non-negligible occupation probability. This is seen in the linear behavior in the log-log plot in the main Fig.~\ref{fig:b0_pvse} where we have used $\beta_0 = 0~, \beta_1= 0.02$ and $\beps = 0.01$. Due to the heavy tailed distribution the excitations are distributed over several energy levels and hence the individual values of $p_{n_+,n_-}$ are very small as shown in Fig. \ref{fig:pmn_b0} within the resonance window.

Outside the PR window, the behavior changes qualitatively. The  distribution decays exponentially,
\begin{equation}
    \frac{P_N}{g_N} \propto e^{-\al_{\text{exp}} N},
    \label{eq:exp}
\end{equation}
as shown in the inset of Fig.~\ref{fig:b0_pvse}, where the same data for a different detuning $\beps = 0.05$ appears linear on a semilog scale. This rapid decay reflects the fact that off-resonant driving fails to sustain excitation into high-$N$  states, and the system is restricted to states close to the ground state.

The crossover between these two regimes therefore serves as a sharp  diagnostic of PR: power-law dependence signals resonant energy pumping into the oscillator, while exponential dependence indicates an off-resonant rapidly decaying response.

\begin{figure}[htb]
    \centering 
    \ig[width=0.82\linewidth]{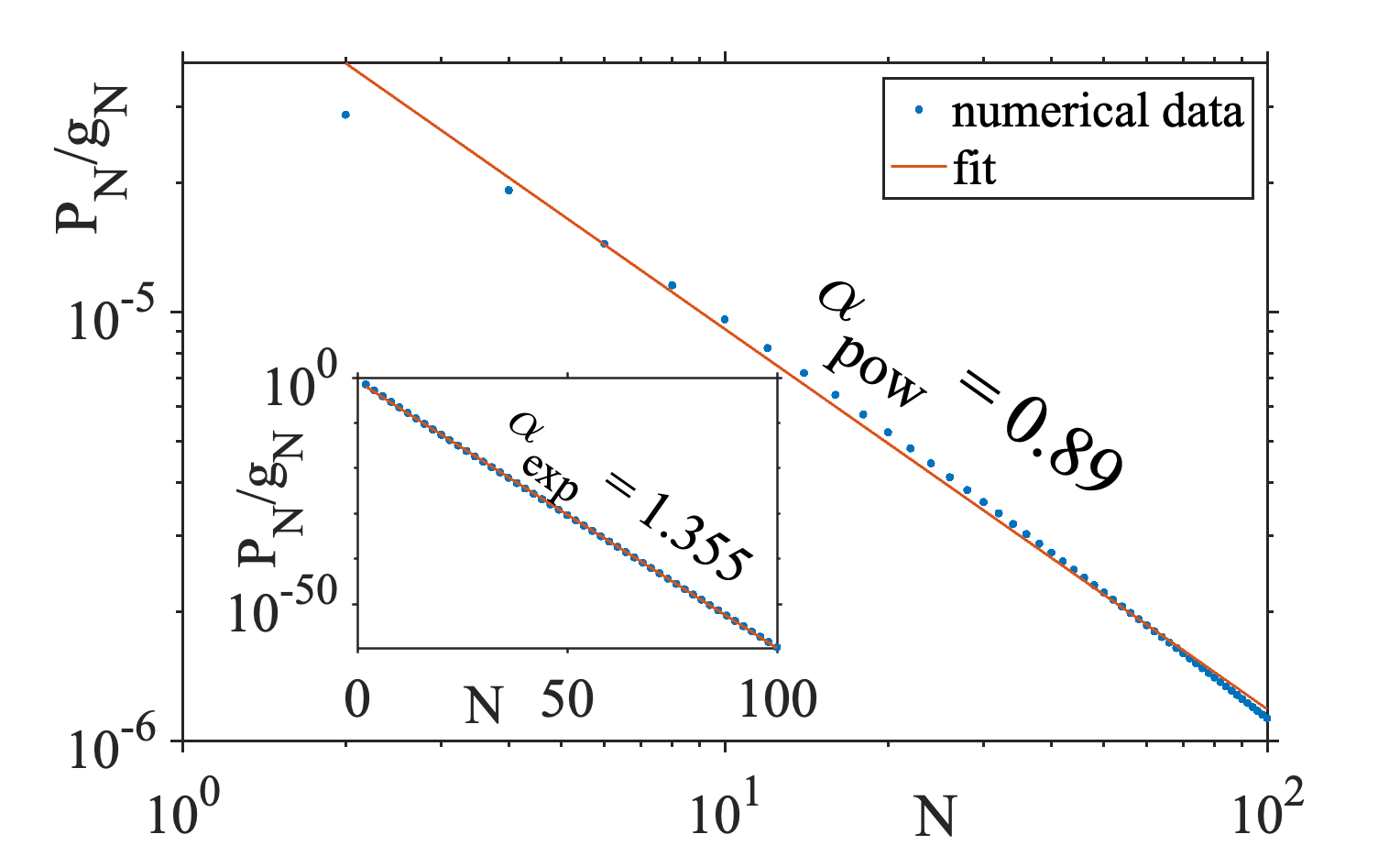}
    \caption{Average probability per state, $P_N/g_N$, as a function of the occupation number $N$. In the PR regime, the distribution follows a power law with  $\al_{\text{pow}} \approx 0.89$, appearing as a straight line on the log-log plot. Outside the PR regime, the behavior becomes exponential, as shown in the inset, where it appears linear on a semilog scale with $\al_{\text{exp}} \approx 1.36$.{Note that both these exhibit decaying behavior due to the minus sign appearing with the exponent in Eqs. \eqref{eq:exp} and \eqref{eq:pow}.}}
    \label{fig:b0_pvse}
\end{figure}

In order to understand the $\beta_0\neq0$ case, we require a slightly different approach. Unlike the $\beta_0=0$ case, the total excitation number is no longer a good quantum number as the energy is no longer proportional to $N$. Instead, it is now given by Eq. \eqref{eq:En}. Therefore, we plot the marginal probabilities defined by 
\beq
p_{n_\pm} = \sum_{n_\mp}p_{n_+,n_-},
\label{eq:marg}
\eeq
i.e. we find the probability of a given normal mode being in a particular level, tracing over the states of the other mode. As can be easily inferred from Fig.~\ref{fig:trend_b1}(a), when the drive frequency is tuned to the PR window of the $+$ mode, $p_{n_+}$ exhibits a power-law decay while $p_{n_-}$ decays exponentially. When the drive is tuned to the PR window corresponding to the $-$ mode then the behavior is flipped, i.e., $p_{n_-}$ exhibits a power-law decay while $p_{n_+}$ decays exponentially as shown in Fig.~\ref{fig:trend_b1}(b). This directly demonstrates the mode-selective nature of the excitation for $\beta_0 \neq 0$.

\begin{figure}[htb]
    \centering
    \ig[width=0.7\linewidth]{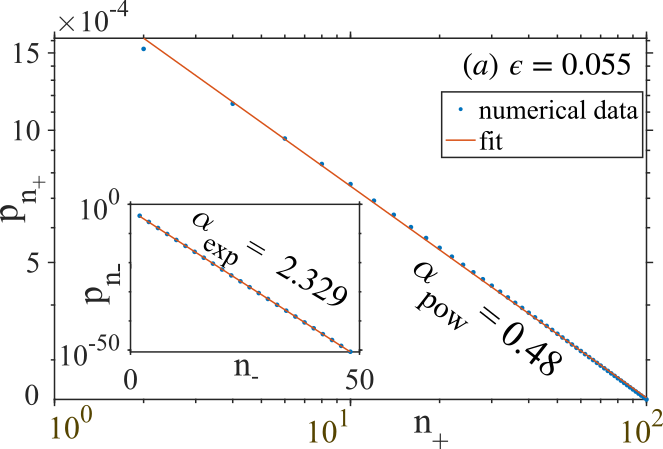}\\
    \ig[width=0.75\linewidth]{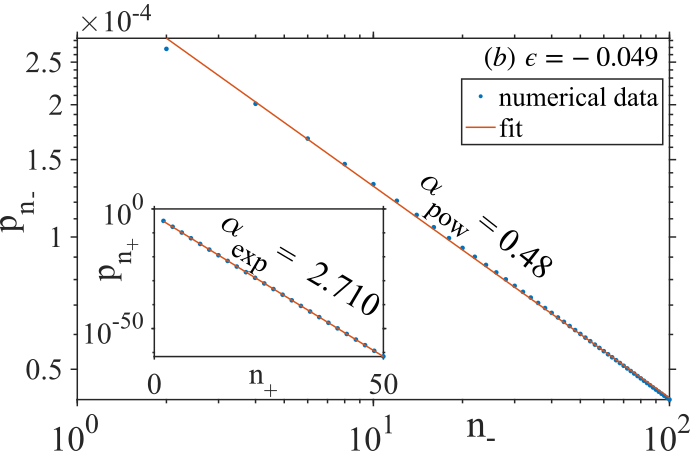}
    \caption{Marginal probabilities defined in Eq. \eqref{eq:marg} for two values of the detuning parameter (a) $\beps = +0.055$ and (b)$\beps = -0.049$. These two values of $\beps$ lie in the two PR windows (troughs) that appear in Fig. \ref{fig:pmn_b1}(b). In (a) we see that the marginal distribution $p_{n_+}$ exhibits a power law behavior while $p_{n_-}$ decays exponentially. This shows that while the $+$ mode is excited to the higher energy states, the $-$ mode continues to remain close to the ground state. On the other hand, these trends are inverted in (b) in the other PR window.}
    \label{fig:trend_b1}
\end{figure}

\subsection{Energy Pumping}

The remarkable feature of resonance is that when the drive frequency closely matches the resonance frequency, the energy pumping into the oscillator is maximized. The energy of the system can be evaluated by computing the expectation value in the normalized time-evolved state given in  Eq.\eqref{eq:psitau}, 
\begin{figure}[htb]
    \centering
    \ig[width=0.9\linewidth]{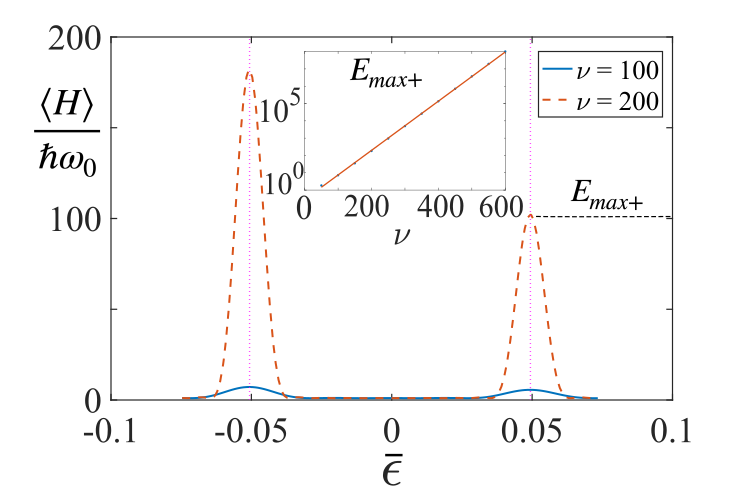}
    \caption{The energy absorbed by the parametrically driven coupled oscillator for a range of $\eps$s. The blue (solid) and red (dotted) curves correspond to two different number of drive cycles $\nu = 100$ and $200$, respectively. The other parameters $\beta_0 (=0.05)$ and $\beta_1 (=0.02)$ are constant.  The energy absorbed peaks in the two resonance windows centered around $\Omega_{\pm0} = 1\pm\beta_0$, consistent with the PR regimes shown in Fig. \ref{fig:pmn_b1}. Depending on the drive frequency, either of the two normal modes is excited. The inset is a semilog plot of the energy peak at $\Omega_{-0}$ as a function of number of drive cycles $\nu$.} 
    \label{fig:energy}
\end{figure}

\begin{eqnarray}
\frac{\expval{H}}{\hbar\omega_0}
&=& 
{\expval{\hat H_+}{\Psi_+}}+ {\expval{\hat H_-}{\Psi_-}}\non \\
&=& 
\frac{\Om_{+0}^2 + 4 |B_{+}|^2}{8B_{+,R}} + \frac{\Om_{-0}^2 + 4 |B_{-}|^2}{8B_{-,R}}
\end{eqnarray}
with $B_\sigma = B_{\sigma,R} + iB_{\sigma,I}$, which is a function of the drive parameters $\beta_0$, $\beta_1$ and $\beps$, is numerically evaluated from Eq. \ref{eq:diff}.

This energy $\langle H \rangle$ in units of $\hbar\omega_0$ as a function of the detuning parameter $\beps$ is shown in Fig.~\ref{fig:energy} for two values of the number of drive cycles $\nu$. We have chosen $\beta_0 = 0.05$ and $\beta_1=0.02$. The energy absorbed peaks sharply at two values of $\eps$ corresponding to the two normal-mode resonance windows $\Omega_{\pm 0} = 1 \pm 
\beta_0$, confirming that the drive selectively pumps energy into one mode at a time. The inset shows that within the resonance window the energy grows exponentially with the number of drive cycles $\nu$, which is the hallmark of parametric instability.

\section{Generalization to N-coupled oscillators} \label{sec:Nosc}

While we have discussed the case of only two coupled oscillators, this treatment can easily be generalized to any number of identical oscillators with a general quadratic coupling. The general Hamiltonian is written compactly in the matrix form as,

\beq
\Hat{H} = \frac{\hat{\mathbf{p}}^2}{2m} + \frac{1}{2}\mathbf{x}^T \mK(t)\, \mathbf{x},
\label{H_mat}
\eeq
with $\mathbf{x} = (x_1, x_2, \ldots, x_N)^T$ and $\hat{\mathbf{p}} = (\hat{p}_1, \hat{p}_2, \ldots, \hat{p}_N)^T$ being the {generalized degrees of freedom and their corresponding generalized momenta, respectively.}

The symmetric matrix $\mK(t)$ can be separated into the static and dynamical components,
\bea
\mK(t) = \begin{cases} \mK_0 + \mK_1 \sin\left[(2\omega_0 +\epsilon)t\right]  ~~~~~ \text{if } 0 < t < t_f ,\\ \mK_0,~~~~~~~~\text{otherwise.}
\end{cases}
\eea
such that the diagonal entries of $\mK_0$ correspond to the natural frequencies of the oscillators, i.e. $\mK^{ii}_{0} = m\omega_0^2$ and the off-diagonal entries $\mK^{ij}_0$ are the static components of the inter-oscillator coupling. The amplitude of parametric modulation is encoded in $\mK_1$ as off-diagonal elements. From this general form given here we can recover Eq. \eqref{H_xy} for the two oscillators case by using the following forms of the coupling matrices,
\bea
\mK_0 = \begin{pmatrix} m\om_0^2 & \kappa_0 \\ \kappa_0 & m\om_0^2\end{pmatrix},~~~ \mK_1 = \begin{pmatrix} 0 & \kappa_1 \\ \kappa_1 & 0\end{pmatrix}.
\eea

Once again, just like the two-oscillator case we can obtain the normal modes by diagonalizing the static coupling matrix $\mathbf{K}_0$. In this normal-mode basis, the Hamiltonian separates into $N$ independent parametrically driven oscillators. The results derived for the two-oscillator case can directly be carried over. The Gaussian ansatz remains valid due to the quadratic Hamiltonian. The selection rule of only the even numbered states being excited holds for each of the $N$ normal modes. 

Finally, the mode-selectivity of the excitations, which is the central result of this work, also generalizes naturally. When the normal mode frequencies are sufficiently well separated, each of the $N$ normal modes has a distinct resonance window. By tuning the drive frequency appropriately, we can selectively excite a single mode while leaving all the other modes close to their respective ground states. 

\section{Conclusions and Outlook}\label{sec:disc}
Parametric resonance applied to quantum systems results in certain features which do not have a classical analogue. This work demonstrates one such phenomenon in which the normal modes in the quantum ground state of a coupled harmonic oscillator can be selectively excited by parametrically modulating the coupling between the degrees of freedom, which is different from the commonly used method of modulating the natural frequency directly. Under specific circumstances, namely, a non-vanishing static component of the coupling parameter, this protocol naturally creates two distinct parametric resonance windows. This means that by suitably tuning the drive frequency to one of these windows, in one of the normal modes several excited states are populated,  while the other normal mode remains close to its ground state.

Additionally, we have established the selection rules for the allowed transitions under such a drive. First of all, we see that only the even numbered states are populated in each of the normal modes, i.e., the system climbs the excitation ladder two rungs at a time. This is a direct consequence of the even parity of the Gaussian wavefunction which forbids the transition to states with odd parity. {A remarkable result of our analysis is that} the PR and non-PR regimes can be distinguished by analyzing the dependence of the probabilities on the energy levels -  a power-law decay within the PR regime in contrast to the exponential decay observed off-resonance. This serves as a sharp diagnostic tool for parametric resonance of quantum systems. All these results carry over to $N$ coupled oscillators as discussed in Sec.~\ref{sec:Nosc}.

There are several natural directions for future work. The most immediate is to study the effect of dissipation. Any realistic physical system, is coupled to an environment or a bath which can lead to relaxation and dephasing. This interplay between the drive and dissipation is known to lead to rich physics even in the single-mode case~\cite{gosner20, kryu96, ferrari19, Foot25}. We expect the two-mode (or, more generally, the $N-$mode) system to show additional structure due to the mode-selective nature. In particular, it would be interesting to examine whether the mode selectivity survives in the presence of dissipation, and how the power-law decay of the excitation spectrum in the PR regime gets modified.

Another natural extension is to consider the effect of a finite switch-on and switch-off time for the drive. In the present work, the drive is assumed to turn on and off instantaneously. In any realistic experimental setting, however, the drive amplitude ramps up and down over a finite timescale. Since the Hamiltonian remains quadratic throughout, the Gaussian ansatz and the resulting selection rules continue to hold for any smooth switching profile. It would be interesting to examine how the width and sharpness of the parametric resonance window, as well as the mode-selectivity condition, are affected as the switching timescale is varied between the sudden and adiabatic limits.

The mode selectivity demonstrated here has potential applications in several quantum systems, including superconducting quantum circuits, optomechanical arrays, and trapped ion systems.

\bibliography{refs}
\end{document}